\documentclass{PoS}

\def\d{\mbox{d}}
\def\Im{\mbox{Im}}
\def\Re{\mbox{Re}}

\title{Unstable-particles pair production in modified
perturbation theory in NNLO}

\ShortTitle{Unstable-particles pair production in modified
perturbation theory in NNLO}

\author{\speaker{Maksim Nekrasov}%
         \\
        Institute for High Energy Physics\\
        E-mail: \email{maksim.nekrasov@ihep.ru}}

\abstract{ We consider pair production and decay of fundamental
unstable particles in the framework of a modified perturbation
theory (MPT), which treats resonant contributions of unstable
particles in the sense of distributions. The cross-section of the
process is calculated within the NNLO of the MPT in a model that
admits exact solution. Universal massless-particles contributions
are taken into consideration. The calculations are carried out by
means of FORTRAN code with double precision, which ensures per
mille accuracy of computations. A comparison of the outcomes with
the exact solution reveals an excellent convergence of the MPT
series at the energies close to and above the maximum of the
cross-section. Near the maximum of the cross-section the
discrepancy of the NNLO approximation makes up a few per mille. }

\FullConference{13th International Workshop on Advanced Computing
and Analysis Techniques in Physics Research\\
February 22-27, 2010\\
Jaipur, India}

\begin{document}

A description of processes of production and decays of fundamental
unstable particles for colliders subsequent to LHC should be made
generally with the NNLO accuracy. This implies that the
calculations must provide gauge cancellations and unitarity, and
enough high accuracy of computation of resonant contributions. The
existing methods, such as the DPA successfully applied at LEP2
\cite{DPA} or the complex-mass scheme (CMS) \cite{CMS} intended
mainly to ILC \cite{ILC}, provide only the NLO precision of the
description. The pinch-technique method \cite{pinch}, another for
a long time developed approach, in principle can provide the NNLO
precision, but for the maintenance of the gauge cancellations it
requires a lot of calculations of extra contributions that become
apparent formally at the next level of the precision, which is
impractical \cite{Ditt}. So alternative approaches are required. A
modified perturbation theory (MPT) \cite{F1,N1,N2} is such an
approach. Its main feature is the direct expansion of the
probability instead of amplitude in powers of the coupling
constant with the aid of distribution-theory methods. As the
expansion is made in powers of the coupling constant and the
object to be expanded is gauge invariant, the gauge cancellations
in the MPT should be automatically maintained. However, the
accuracy of the description remains unknown. To clear up this
question numerical simulations are required. In this report I
present results of the appropriate calculations and,
simultaneously, do a brief introduction to the MPT method.

Actually I will discuss the case of pair production of unstable
particles. The most crucial in this case are the double-resonant
contributions. The corresponding total cross-section, for example
in $e^+ e^-$ annihilation, has the form of a convolution of the
hard-scattering cross-section with the flux function,
\begin{equation}\label{not1}
\sigma (s) = \int_{s_{\mbox{\tiny min}}}^s \frac{\d s'}{s} \:
\phi(s'/s;s) \> \hat\sigma(s')\,.
\end{equation}
(In fact the angular distributions may be described in the MPT, as
well, but I do not discuss this option below.) The hard-scattering
cross-section has the form of an integral over the virtualities of
unstable particles,
\begin{equation}\label{not2}
\hat\sigma (s) =
 \int \!\!\!\! \int \d s_1 \, \d s_2 \;
 \theta(\!\sqrt{s}-\!\sqrt{s_1}-\!\sqrt{s_2}\,)
 \sqrt{\lambda (s,s_{1},s_{2})}\;\Phi(s;s_1,s_2)
 \left(1\!+\!\delta_{c}\right) \,
 \rho(s_{1}) \> \rho(s_{2})\,.
\end{equation}
Here the first two multipliers in the integrand are kinematic
factors, $\rho(s_i)$ are Breit-Wigner (B$\!$W) factors,
$(1\!+\!\delta_{c})$ stands for soft massless-particles
contributions, and function $\Phi$ is the rest of the amplitude
squared. Generally $\Phi$ corresponds to one-particle irreducible
contributions, and hence $\Phi$ has no singularities on the
mass-shell of unstable particles. On the contrary, the kinematic
factors have singularities due to the theta function and the
square root of the kinematic function $\lambda$. The B$\!$W
factors if to naively expand them in powers of the coupling
constant $\alpha$ generate nonintegrable singularities. Actually
this constitutes a serious problem since integrals in (\ref{not2})
become senseless.

The singularities, nevertheless, become integrable if to expand
the B$\!$W factors in the sense of distributions. In this case the
expansion of a separately taken B$\!$W factor is beginning with
the $\delta$-function which corresponds to the narrow-width
approximation. The contributions of the naive Taylor expansion are
supplied with the principal-value prescription for the poles. The
nontrivial contributions are the delta-function and its
derivatives with coefficients $c_{n}$, which are polynomials in
$\alpha$ determined by the self-energy of the unstable particle
\cite{F1}. Within the NNLO, the expansion has the form
\begin{eqnarray}\label{not3}
&\displaystyle \rho(s) \;\;\equiv\;\; \frac{M\Gamma_0}{\pi} \; {|s
- M^2 + \Sigma(s)|^{-2}} \;\;=&
\\
&\displaystyle \quad\;\;
    \delta(s\!-\!M^2) \,+\,
    \frac{M \Gamma_{0}}{\pi} \, PV \! \left[\,\frac{1}{(s-M^2)^2}
  -\, \frac{2\alpha\,\Re\Sigma_1(s)}{(s\!-\!M^2)^3}\,\right] \;+\;
  \sum\limits_{n\,=\,0}^2 c_{n}(\alpha)\,
  \frac{\mbox{\small ($-$)}^{n}}{n!}\,\delta^{(n)}(s\!-\!M^2) +
  O(\alpha^3)\,.&\nonumber
\end{eqnarray}
Here $M$ is the renormalized mass, $\Gamma_{0}$ is the Born width,
$\Sigma(s)$ is the self-energy of the unstable particle.
Coefficients $c_n$ within the NNLO include 3-loop self-energy
contributions and their derivatives, determined on-shell. The
structure of the contributions is such that in the OMS-type
schemes of the UV renormalization the real self-energy
contributions enter into the coefficients either without the
derivatives or with the first derivative only. This implies that
the relevant real self-energy contributions are determined by the
renormalization conditions. Unfortunately, the conventional OMS
scheme is not convenient, since it does not maintain the gauge
independence of the renormalized masses of unstable particles. So
it is reasonable to proceed to the particular version of the OMS
scheme, namely to the $\overline{\mbox{OMS}}$ or pole scheme
\cite{OMS-bar,Sirlin}, where the renormalized masses of unstable
particles by definition coincide with the real parts of the poles
of the propagators and thereby coincide with the observable
masses. The coefficients $c_n$ in this scheme are determined as
follows \cite{N2}:
\begin{equation}\label{not4}
c_0  =  - \, \alpha \, \frac{I_2}{I_1} + \alpha^2
\left[\frac{I_2^2}{I^2_1} - \frac{I_3}{I_1} - (I_{1}^{\,\prime})^2
\right],\qquad c_1 = 0, \qquad c_2 = -\, \alpha^2 I^2_1 \,.
\end{equation}
Here $I_{k} = \Im\,\Sigma_{k}(M^2)$, $\Sigma = \alpha \,\Sigma_1 +
\alpha^2 \,\Sigma_2 + \alpha^3 \,\Sigma_3$, and $I_{1}^{\,\prime}
= \Im\,\Sigma^{\,\prime}_{1}(M^2)$. Simultaneously in the
$\overline{\mbox{OMS}}$ scheme the $\Im\,\Sigma(M^2)$ coincides
with the imaginary part of the pole of the propagator. This allows
one to connect $I_{k}$ order-by-order with the width of the
unstable particle via the unitarity relations $\alpha I_1 =
M\Gamma_0$, \ $\alpha^2 I_2 = M \alpha \Gamma_1$, \ and \
$\alpha^3 I_3 = M \alpha^2 \Gamma_2 +
\Gamma_0^3/(8M)$~\cite{OMS-bar}.

In fact, expansion (\ref{not3}) has sense only if the weight in
the integral is a regular enough function. In our case, however,
the kinematic factor is not regular, and this leads to the
divergence of integrals (\ref{not2}) after the substitution of the
expansions. At first glance this closes a possibility of
application of the expansions for the B$\!$W factors.
Nevertheless, the kinematic factor may be analytically regularized
via the substitution $[\lambda (s,s_{1},s_{2})]^{1/2} \to [\lambda
(s,s_{1},s_{2})]^{\nu}$. With large enough $\lambda$ this imparts
enough smoothness to the weight, and the singular integrals become
integrable. Moreover, after the calculation of the integrals and
removing the regularization the outcomes remain finite and the
expansion remains asymptotic \cite{N2}. In principle, this
salvages the applicability of the approach.

However, the singular integrals must be analytically calculated.
The scheme of their calculation is as follows. At first one should
proceed to dimensionless energy variables $x$, $x_i$ ({\small {\it
i} = 1,2}) counted off from thresholds, $\sqrt{s} = 2 M (1 +
x/4)$, $\sqrt{s_{i}} = M(1 + x_{i}/2)$. Formula (\ref{not2}) for
the hard-scattering cross-section (with the analytically
regularized kinematic factor) then takes the form
\begin{equation}\label{not5}
 \widetilde{\sigma}(x) = \int\!\!\!\!\int\!\d x_1 \, \d x_2 \;
 (x\!-\!x_1\!-\!x_2)_{+}^{\nu} \;
 \widetilde{\Phi}(x\,;x_1,x_2) \;
 \widetilde{\rho}(x_1) \widetilde{\rho}(x_2)\,.
\end{equation}
Here $(x\!-\!x_1\!-\!x_2)_{+}^{\nu} = \theta(x\!-\!x_1\!-\!x_2)
(x\!-\!x_1\!-\!x_2)^{\nu}$ and tilde marks the dimensionless
functions. (For convenience, factor $(1\!+\!\delta_{c})$ is
included into the definition of the test function
$\widetilde{\Phi}$.) Further, I substitute asymptotic expansions
for $\widetilde{\rho}(x_i)$ and consider at every $n_i$ in the
product of the expansions the contributions of the $PV x_i^{-n_i}$
and $\delta^{(n_{i}-1)}(x_i)$. Simultaneously, in each case, I
represent the test function in the form of double Taylor expansion
truncated at $x_1^{(n_1-1)}$ and $x_2^{(n_2-1)}$ plus a remainder,
\begin{equation}\label{not6}
{\widetilde{\Phi}(x\,;x_1,x_2)} \; = \sum_{k_1=0}^{n_1-1}
\sum_{k_2=0}^{n_2-1}  \frac{x_1^{k_1}}{k_1 !}\;
\frac{x_2^{k_2}}{k_2 !} \:\;
\widetilde{\Phi}^{(k_1,\,k_2)}(x\,;0,0) \;+\;
\Delta\widetilde{\Phi}(x\,;x_1,x_2)\,.
\end{equation}
The higher powers of $x_1$ and $x_2$ in the Taylor expansion will
zero the $\delta^{(n_{i}-1)}(x_i)$ and cancel the $PV x_i^{-n_i}$.
The remainder $\Delta\widetilde{\Phi}$ is determined as the
difference between $\widetilde{\Phi}$ and the Taylor expansion. In
fact $\Delta\widetilde{\Phi}$ is to be further expanded with
respect to separately $x_1$ and $x_2$, but for brevity I do not
consider this procedure explicitly (see details in \cite{N2}). I
mention only that the final remainder produces a regular
contribution to the integrand in formula (\ref{not5}), and the
integrals of this contribution can be numerically calculated. At
the same time, the contributions of the Taylor expansions are
singular, and the integrals of them are analytically calculable.
After the analytic integrating and putting $\nu \to 1/2$, a sum of
regular and singular contributions appears with the singular
contributions being products of regular factors and the power
distributions of the type $x^{5/2\,-\,n}_{+}$ with integer~$n$.

The convolution integral (\ref{not1}) of above contributions can
be numerically calculated. In particular, the integral of singular
power distributions may be calculated by means of the formula
\vspace*{-0.1\baselineskip}
\begin{equation}\label{not7}
\int \d x \;\; x_{+}^{\nu} \,\phi(x) =
 \int\limits_{0}^{\infty} \d x \;\; x^{\nu}
 \left\{\phi(x) -
 \sum_{k=0}^{N-1} \frac{x^{k}}{k!} \, \phi^{(k)}(0)\right\}\,,
\vspace*{-0.2\baselineskip}
\end{equation}
where $\phi$ is a weight and $N$ is a positive integer such that
$-N\!-\!1< \Re \, \nu < -N$. The test function $\widetilde{\Phi}$
at this stage is considered determined in the conventional
perturbation theory.

For carrying out the above mentioned calculations rather general
FORTRAN code with the double precision is written. The calculation
of regular integrals in this code is fulfilled by Simpson method.
Numerous indeterminate forms of the type $0/0$ that emerge in the
integrand due to the difference structures are resolved through
the introduction of linear patches. The patches diminish the
errors that arise because of the loss of decimals near the
indeterminacy points, and manifest themselves, in particular, in
numerical instability. Of course, the patches generate the errors,
too. However the latter errors are under control and are estimated
as $\epsilon^{2} \varphi''_0/(3\varphi_0)$, where $\epsilon$ is
the size of the patch, $\varphi_0$ and $\varphi''_0$ is the
integrand and its second derivative at the center of the patch.
This estimate has sense of a relative error of the calculation of
integral inside the patch. The sizes of the patches are so
determined that the singular components of indeterminate forms
take a constant value on the boundaries of the patches. So, for
instance, in the case of an indeterminate form of the type
$x^n/x^n$ the size $\epsilon$ should be so chosen that
$\epsilon^n$ would be a constant, say, $10^{-N}$ with
some~fixed~$N$.

Unfortunately, the errors caused by the loss of decimals near the
indeterminacy points but outside of the patches, cannot be
explicitly estimated. What is possible to formulate is only the
dependence of these errors on the basic parameters. So, in the
case of indeterminacy $[f(x)-f(0)]/x$ the appropriate error has a
behavior $\sim 10^{-(D-N)} f_0/f^{\,\prime}_0$, where $D$ is the
number of digits in the representation of real numbers (in our
computations $D=15$), and $N$ is the above mentioned parameter of
smallness of $\varepsilon$. In the case of $[f(x)-f(0)-x
f^{\,\prime}(x)]/x^2$, the analogous estimate is $\sim 10^{-(D-N)}
\, f_0/f^{\,\prime\prime}_0$, where $N$ is the order of smallness
of $\varepsilon^2$ etc.

Nevertheless, the total error of the computations can be
estimated. The crucial point is that the errors because of the
patches are increasing with increasing the sizes of the patches,
while the errors because of the loss of decimals are decreasing.
This implies that there should be an optimum size of the patches
when the sum of all the errors is minimized. The point of the
minimization must possess extremum properties, so that the result
of the computation at this point must be stable with respect to
varying the sizes of the patches. Moreover, at the extremum point,
the sums of the errors of different kinds must be approximately
equal each other (up to a coefficient of order one). So the order
of the total error of the computation may be estimated by the
order of the sum of errors because of the patches. However, the
latter sum can be numerically estimated. By this means the total
error of the computations can be estimated. In particular, in the
case of the model discussed below, the optimum size of the patches
is found established by $N \approx 8$ (at $D=15$). The
corresponding estimate of the relative error of the computation of
the NNLO approximation~constitutes~$10^{-3}$.

Now I proceed to the physical model in the framework of which I do
computation. First I recall that my aim is to verify whether the
MPT calculations are practicable in principle, and then my aim is
to test the convergence properties of the MPT expansion. The
existing experience \cite{N3} shows that the convergence
properties are determined mainly by the B$\!$W factors, but not by
particular form of the test function. So without loss of
generality I can determine the test function $\widetilde{\Phi}$ in
the framework of a model. As such a model, I consider the Born
approximation for the process~$e^{+} e^{-} \to \gamma,Z \to t\bar
t \to W^{+} b\:W^{-}\bar b$. Simultaneously, I consider the
self-energies in the denominators of the top quark propagators
with the 3-loop contributions \cite{N3}. Since soft massless
particles generally can make considerable contributions, I take
into consideration also the universal soft massless-particles
contributions. Actually they are collected in the flux function
and in the Coulomb factor. The flux function, I consider in the
leading-log approximation. In the Coulomb factor determined mainly
by gluons, the multi-gluon contributions, generally, are
important. However, at least at a distance from the threshold a
qualitative picture may well be simulated by the one-gluon
contributions. So, in the framework of a model I consider Coulomb
factor in the one-gluon approximation. In addition, I imply the
standard resummation in the Coulomb factor \cite{Coulomb}, which
does not affect the B$\!$W factors.

\begin{figure}
\includegraphics[width=\textwidth]{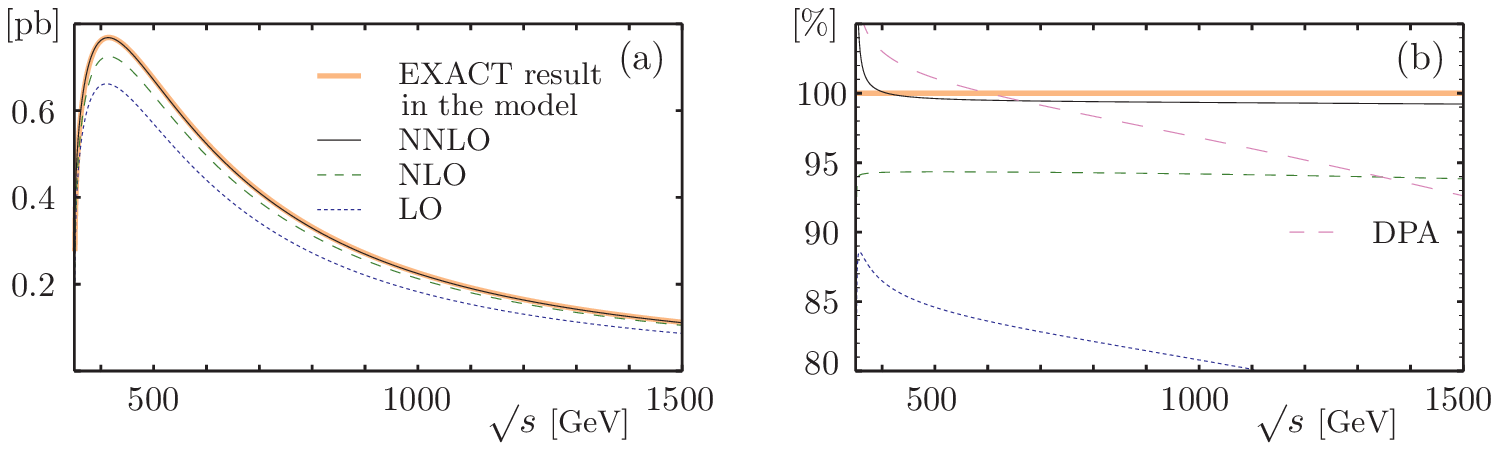}
\caption{The total cross-section: the exact result in the model
and the results of computations in MPT.} \label{fig1}
\end{figure}

\begin{table}[t]
\begin{center}
\begin{tabular}{ c  c c c c  }
\hline\noalign{\medskip} \\[-6mm]
 $\quad \sqrt{s}$ (TeV) $\qquad$
 & $\qquad \sigma_{\,\mbox{\tiny EXACT}} \qquad$      &
 $\qquad \sigma_{LO} \qquad$        & $\qquad \sigma_{NLO} \qquad$ &
 $\quad \sigma_{NNLO}   \qquad$ \\
\hline\noalign{\medskip} \\[-6mm]
 0.5                               & 0.6724         &
 0.5687          &  0.6344         & 0.6698(7)          \\
                                   & {\small 100\%} &
 {\small 84.6\%} & {\small 94.3\%} & {\small 99.6(1)\%} \\
\hline\noalign{\medskip} \\[-6mm]
 1                                 & 0.2255         &
 0.1821          &  0.2124         & 0.2240(2)          \\
                                   & {\small 100\%} &
 {\small 80.8\%} & {\small 94.2\%} & {\small 99.3(1)\%} \\[-1mm]
\noalign{\smallskip}\hline
\end{tabular}
%\caption{\small The total cross-section in pb and in \% with
%respect to exact result in the model.}
\end{center}\vspace*{-8mm}
\end{table}

The outcomes of the computations are presented in the Figure and
in the Table. In the Figure in the panel (a) the thick curve shows
the behavior of the total cross-section in the model. The dotted,
dashed, and continuous thin curves show the results of the MPT
computations in the LO, NLO, and NNLO approximations,
respectively. It is worth noting that the NNLO result almost
coincides with the exact result in the model. The distinction is
visible in the panel (b) where the percentages with respect to the
exact result are presented. In the Table the outcomes are
represented in the numerical form at the characteristic energies
planned at the ILC \cite{ILC}. In the last column the numbers in
parenthesis represent the uncertainties in the last digit. In
other columns the uncertainties are omitted as they appear beyond
the precision of the presentation of data.

In conclusion, first of all the above results show in practice the
existence of the MPT expansion in the case of pair production and
decay of fundamental unstable particles. Secondly, the NLO and
NNLO approximations in the MPT have very stable behavior at the
energies near the maximum of the cross-section and at higher
energies (and this result to a large extent is model-independent
\cite{N3}). Thirdly, at the ILC energies the NNLO approximation in
the MPT in the case of the top-quark pair production provides
approximately a half-percent accuracy of the description of the
cross-section. In fact this is what is needed at the ILC. However,
the higher precision of the MPT description, when needed, may be
achieved by the proceeding to the NNNLO, or by the using of the
NNLO for the calculation of the loop contributions only---on the
analogy of actual practice of application of DPA \cite{DPA}. The
higher precision of the actual computations in the NNLO of MPT may
be achieved at the proceeding to the representation of real
numbers with higher precision (higher than with the double
precision).

\medskip

The author is grateful to A.L.Kataev for encouragement and
suggestion to make a report at the ACAT Workshop.

\end{document}